\documentclass[twocolumn]{aastex}
    
\usepackage[english]{babel}
\usepackage{amsmath}
\usepackage{graphicx}
\usepackage{wasysym}
\usepackage{soul}
\usepackage{parskip}
\usepackage{wrapfig}
\usepackage{url}
\usepackage{float}
\usepackage{hyperref}
\usepackage{xcolor}
\usepackage{bm}

\colorlet{Mycolor1}{green!72!red!228!}

\shorttitle{The Habitable Zones of Rapidly Rotating Main Sequence A/F Stars}
\shortauthors{Ahlers et al.}
\graphicspath{{./}{figures/}}

\begin{document}

\title{The Habitable Zones of Rapidly Rotating Main Sequence A/F Stars}

\correspondingauthor{John P. Ahlers}
\email{johnathon.ahlers@nasa.gov}

\author[0000-0003-2086-7712]{John P. Ahlers}
\altaffiliation{NASA Postdoctoral Program Fellow}
\affiliation{Exoplanets and Stellar Astrophysics Laboratory, Code 667, NASA Goddard Space Flight Center (GSFC), Greenbelt, MD 20771, USA}
\affiliation{GSFC Sellers Exoplanet Environments Collaboration}

\author{Emeline F. Fromont}
\altaffiliation{NASA Intern}
\affiliation{Exoplanets and Stellar Astrophysics Laboratory, Code 667, NASA Goddard Space Flight Center (GSFC), Greenbelt, MD 20771, USA}
\affiliation{Department of Physics, Carnegie Mellon University, Pittsburgh, PA 15213, USA}

\author{Ravi Kopparappu}
\affiliation{GSFC Sellers Exoplanet Environments Collaboration}
\affiliation{NASA Goddard Space Flight Center, Greenbelt, Maryland, USA}

\author{P. Wilson Cauley}
\affiliation{Laboratory for Atmospheric and Space Physics, University of Colorado Boulder, 600 UCB, Boulder, CO 80303, USA}

\author[0000-0003-4346-2611]{Jacob Haqq-Misra}
\affiliation{Blue Marble Space Institute of Science, Seattle, WA, USA}

\begin{abstract}
We investigate how rapid stellar rotation commonly seen in A/F stars can influence planet habitability. Specifically, we model how rapid rotation influences a planet's irradiation and determine the location of the habitable zone for stars in the mass range $1.3M_\odot\leq M_\star\leq 2.2M_\odot$. Rapid stellar rotation can dramatically change a star's luminosity and spectral energy distribution and, therefore, can impact the habitability of any surrounding planets. Stars of mass $M_\star\gtrsim1.3M_\odot$ commonly rotate near their breakup speeds, which causes two effects relevant to planet habitability. First, these stars flatten into oblate spheroids with shorter polar radii and elongated equatorial radii. Second, rapid rotation induces a pole-to-equator temperature gradient on the surface of these stars. Using a 1D climate model, we calculate the inner and outer edges of the habitable zone of well-known rapid rotators and average theoretical stars in our stellar mass range. We find that, in general, rapid rotation causes the habitable zone to reside closer in than for a non-rotating equivalent star. We also find that gravity darkening dramatically reduces stellar UV emission, which combats the common assumption that high-mass stars emit too much UV light for habitable worlds. Overall, we determine that rapid stellar rotation has important consequences for the overall habitability of a system and must be accounted for both when modeling exoplanet environments and in observation of planets around high-mass stars. 
\end{abstract}

\section{Introduction}\label{sec:intro}
Several recent works have included F-stars and even A-stars in the search for habitable worlds. For instance, \citet{ramirez2018methane} argue that because life on Earth arose at least as quickly as $\sim700$ Myr after its formation \citep{mojzsis1996evidence}, stars whose main sequence lifetimes last at least 700 Myr should be included in habitable zone (HZ) definitions. Therefore, in the spirit of open-mindedness, recent HZ boundary calculations have included A-stars with $T_\mathrm{eff}$ up to $\sim10,000$\,K \citep[e.g.,][]{danchi2013effect,ramirez2018methane,ramirez2018more}. Stars in this mass range commonly rotate rapidly throughout their main sequence lifetimes; in this publication, we investigate the impacts of rapid stellar rotation on the location of the HZ and overall planet habitability.

The high-mass stars we consider ($1.3M_\odot\leq M_\star\leq 2.2M_\odot$, $6300\,\mathrm{K}\leq T_\mathrm{eff}\leq10000\,\mathrm{K}$) make interesting targets for hosting potentially habitable worlds. Stars in this mass range emit a high amount of short wavelength light that recent works predict to be necessary for biologic genesis and evolution. \citet{haqq2017drake} and \citet{haqq2019does} explored planet evolutionary timescales as a function of a host star's spectral energy distribution and found that F- and G-dwarf stars are the most likely places for life to currently exist. Specifically, stars that emit a high fraction of light between 200\,nm and 1200
\,nm are well-suited for generating life. Photons in this wavelength range can benefit life by driving genetic mutation and contributing to abiogenesis.

Recent studies have shown that a Goldilocks abundance of UV light produces favorable conditions for habitability: enough to trigger biogenesis, but not so much as to harm biologic processes such as DNA recombination. \citet{haqq2019does} identifies UV radiation above 200 nm to be beneficial for producing beneficial mutations and shortening evolution timescales. \citet{buccino2007uv} and others \citep[e.g.,][]{guo2010habitable,oishi2016simple} introduced the concept of the UV habitable zone (UVHZ), which is the distance from a star that a planet receives UV abundance conducive to generating life. They found that stars with $T_\mathrm{eff}$ between $4600$ K and $7200$ K produce strong HZ and UVHZ overlap, corresponding to favorable conditions in the mass range $0.78M_\odot\leq M_\star\leq1.75M_\odot$ at zero age main sequence and $0.59M_\odot\leq M_\star\leq1.91M_\odot$ at the terminal of main sequence. However, these works did not account for rapid stellar rotation, which we show in this work can drastically reduce a star's UV output. Therefore, it is possible that the HZ and UVHZ can overlap at temperatures higher than $7200$ K, suggesting that habitability should be considered for high-mass stars.

Stars above the Kraft break ($M_\star\geq 1.3 M_\odot$) commonly rotate near their breakup speeds \citep{kraft1967studies}, which can cause orbiting exoplanets to receive unusual yearly instellation patterns -- particularly in the NUV and optical wavelengths. \citep{ahlers2016gravity}. Rapid stellar rotation produces two significant effects on a planet's instellation. First, the high centrifugal force at the stars' equators causes them to flatten into oblate spheroids. For example, the rapidly-rotating A7V star Altair spins at about 71\% of its breakup speed and has an equatorial radius $25$\% larger than its polar radius \citep{monnier2007imaging}. Second, their high rotation rate adjusts the stars' effective temperature, producing a pole-to-equator temperature gradient known as gravity-darkening. Altair's effective temperature varies from 6900 K to 8500 K across its surface, resulting in poles 2.3 times more luminous than its equator. Additionally, recent observations show that exoplanets orbiting early-type stars (hereafter called early-type exoplanets) commonly misalign from their host stars' equatorial planes \citep[e.g.,][]{winn2010hot,dawson2014tidal,albrecht2021preponderance}. \citet{ahlers2016gravity} showed that a planet's equilibrium temperature can vary dramatically throughout its orbit due to its varying exposure to the star's hotter poles and cooler equator. Such an effect likely drives radiative forcing on an Earth-like planet, directly impacting its sea surface temperature and hydrological cycle \citep[e.g.,][]{forster2007changes}. 

In this paper, we show how rapid stellar rotation affects the location of the habitable zone. Our simulations primarily focus on a hypothetical Earth-like planet around the well-known rapid rotator Altair, but we include other observed rapid rotators and theoretical average rapid rotators from the mass range $1.35M_\odot-2.2M_\odot$ to test this effect across the spectrum of A/F stars. In \S\ref{sec:methods} we detail our approach to modelling rapidly rotating stars and exoplanet climatic response to varying instellation. In \S\ref{sec:results} we show that gravity darkening can significantly impact the location of the habitable zone and the planet's irradiance. In \S\ref{sec:discussion} we discuss how rapid stellar rotation can affect planet habitability and explore future avenues of investigating this scenario.

\section{Methods}\label{sec:methods}
The two goals of this project are to model rapid stellar rotation's effect on habitable zone locations and to explore how rapid rotation can cause a planet's instellation to vary over time. We model the planets' received irradiation following \citet{ahlers2016gravity}, incorporating the rotating star model from \citet{lara2011gravity}. Using these instellation values, we calculate the inner and outer edges of the habitable zone for an Earth-like planet using a 1-D climate model \citep{kopparapu2013habitable,kopparapu2014habitable}. 

Every scenario we model follows this general path:
\begin{enumerate}
\item Model the host star's spectral energy distribution, accounting for rapid rotation.
\item Calculate the planet's instellation throughout its orbit.
\item Model the planet's climatic response.
\end{enumerate}

We detail each step of this approach in the following subsections.

\subsection{Stellar Model}\label{sec:stellarmodel}
We model the spectral energy distributions (SEDs) of A/F stars using the Roche model of a rotating star \citep{lara2011gravity}. Rapid rotation causes two stellar effects that influence an orbiting planet's irradiation. First, rapid rotators flatten into oblate shapes due to the high centrifugal force at their equatorial regions. Second, rapid rotation produces a pole-to-equator temperature gradient that can cause a star's poles to be more than two thousand Kelvin hotter than its equator. Together, these effects, commonly known as gravity darkening, can dramatically impact a planet's instellation.

A planet's location in its system determines its exposure to its host star's gravity-darkening effect. A planet residing in the star's equatorial plane receives most of its light from the star's cooler, dimmer equator, whereas a planet located above/below one of the host star's poles is primarily irradiated by the star's hotter, brighter pole. Additionally, the planet's location relative to the stellar poles affects the star's solid angle. When above a stellar pole, the planet sees a projected disk at the size by the star's equatorial radius. When in the equatorial plane, the planet sees a projected ellipse with semi-major and semi-minor axes at the star's polar and equatorial radii, respectively. Therefore, a planet above a stellar pole sees a star that is both brighter and larger than when in the equatorial plane. A planet in an inclined orbit oscillates between these two scenarios at twice its orbital frequency.

We model stellar gravity darkening following \citet{lara2011gravity}, which models stellar flattening and effective temperature variation based on on the Roche model 
\begin{equation}
    \frac{GM}{r}+\frac{1}{2}\Omega^2r^2\sin^2(\theta)=\mathrm{const},
\end{equation}
where $\theta$ is the star's colatitude, $r\equiv r(\theta)$ is the stellar radius at any point on the surface, and $\Omega$ is the rotation rate. A/F stars of type F6V $(M_\star\gtrsim1.3M_\odot)$ and earlier typically rotate at a significant fraction of their breakup speeds. They can flatten dramatically, with equatorial radii up to  50\% larger than their polar radii and equatorial effective temperatures more than two thousand Kelvin cooler than their polar temperatures. We model these effects as a function of the fractional breakup speed $\omega\equiv\Omega_\star/\Omega_\mathrm{br}$. We assume rigid surface rotation for all stars, which previous works have determined to be a safe approximation in this stellar mass range \citep{lara2011gravity,bouchaud2020realistic}.

We model the SEDs of well-known rapidly rotating A/F stars (Table \ref{table:famousstars}) to determine how their fast rotation rates can impact their habitable zones. We also model theoretical rapid rotators ranging from $1.35M_\odot\leq M_\star\leq2.2M_\odot$ (Table \ref{table:theoreticalstars}) using average rotational velocity values from \citet{wolff1997angular} to establish the importance of gravity darkening to planet habitability as a function of stellar mass.

\begin{table}
\begin{tabular}{l c c} \hline \hline
    {\bf Parameter} & {\bf Model Value} & {\bf Published Value} \\ \hline
     \multicolumn{3}{c}{Altair \citep{monnier2007imaging}} \\
     $M_\star$ ($M_\odot$) &$1.791$ & $1.791\pm 0.018$\\
     $T_\mathrm{pole}$ (K) &$8450$ & $8450\pm140$ \\
     $T_\mathrm{eq}$ (K) &$6855$ & $6860\pm150$ \\
     $R_\mathrm{pole}$ ($R_\odot$) &$1.634$ & $1.634\pm 0.011$\\
     $R_\mathrm{eq}$ ($R_\odot$) &$2.03661$ & $2.029\pm 
     0.007$\\
     $\omega$ & $0.926795$& $0.923\pm0.006$\\ \hline
     \multicolumn{3}{c}{Vega \citep{monnier2012resolving}} \\
     $M_\star$ ($M_\odot$) & $2.15$ & $2.15^{+0.10}_{-0.15}$ \\
     $T_\mathrm{pole}$ (K) &$10070$ & $10070\pm90$\\
     $T_\mathrm{eq}$ (K) &$9106$ & $8910\pm130$\\
     $R_\mathrm{pole}$ ($R_\odot$) &$2.418$ & $2.418\pm 0.012$\\
     $R_\mathrm{eq}$ ($R_\odot$) &$2.66213$ & $2.726\pm 0.006$\\
     $\omega$ &$0.774$ & $0.774\pm0.012$\\ \hline
     \multicolumn{3}{c}{Kelt-9 \citep{gaudi2017giant,ahlers2020kelt}} \\
     $M_\star$ ($M_\odot$) &$2.52$ &$2.52^{+0.25}_{-0.20}$ \\
     $T_\mathrm{pole}$ (K) & $10170$& $10170\pm 450$\\
     $T_\mathrm{eq}$ (K) & $9541$& $9400\pm 500$\\
     $R_\mathrm{pole}$ ($R_\odot$) &$2.17729$ & $2.17729\pm 0.02733$\\
     $R_\mathrm{eq}$ ($R_\odot$) &$2.39$ &$2.39\pm 0.03$ \\
     $\omega$ &$0.645961$ & $0.645961^{+0.101095}_{-0.126369}$\\ \hline
     \multicolumn{3}{c}{$\alpha$ Cephei \citep{zhao2009imaging}} \\
     $M_\star$ ($M_\odot$) &$1.92$ &$1.92\pm0.04$ \\
     $T_\mathrm{pole}$ (K) &$8588$ & $8588\pm 300$\\
     $T_\mathrm{eq}$ (K) &$7155$ & $ 6574\pm200$\\
     $R_\mathrm{pole}$ ($R_\odot$) &$2.162$ & $ 2.162\pm 0.036$\\
     $R_\mathrm{eq}$ ($R_\odot$) &$2.73815$ & $ 2.740\pm 0.044$\\
     $\omega$ &$0.941$ & $0.941\pm 0.020$\\ \hline
     \multicolumn{3}{c}{$\alpha$ Ophiuchi \citep{zhao2009imaging}} \\
     $M_\star$ ($M_\odot$) &$2.10$ &$2.10\pm0.02$ \\
     $T_\mathrm{pole}$ (K) &$9300$ & $9300\pm 150$\\
     $T_\mathrm{eq}$ (K) &$8059$ & $7460\pm 100$\\
     $R_\mathrm{pole}$ ($R_\odot$) &$2.390$ & $ 2.390\pm 0.014$\\
     $R_\mathrm{eq}$ ($R_\odot$) &$2.87042$ & $ 2.871\pm 0.020$\\
     $\omega$ &$0.885$ & $0.885\pm 0.011$\\ \hline
     \multicolumn{3}{c}{$\beta$ Cassiopeiae 
     \citep{che2011colder}} \\
     $M_\star$ ($M_\odot$) &$1.91$ &$1.91\pm0.02$ \\
     $T_\mathrm{pole}$ (K) &$7208$ & $7208^{+42}_{-24}$
     \\
     $T_\mathrm{eq}$ (K) & $6173$&  $6167^{+36}_{-21}$\\
     $R_\mathrm{pole}$ ($R_\odot$) &$3.06$ &$3.06^{+0.08}_{-0.07}$ \\
     $R_\mathrm{eq}$ ($R_\odot$) &$3.78771$ & $3.79^{+0.10}_{-0.09}$\\
     $\omega$ & $ 0.920$& $0.920^{+0.024}_{-0.034}$\\ \hline
\end{tabular}
\caption{Model parameters of six well-known rapid rotators. We use observed values for mass, polar temperature, polar radius, and rotational velocity for each star as inputs to our gravity darkening model. We generate SEDs for each star (\S\ref{sec:stellarmodel}) to calculate an orbiting planet's irradiance as a function of position (\S\ref{sec:irradiance}).}
\label{table:famousstars}
\end{table}

\begin{table*}
\centering
\begin{tabular}{l l l l l l l l l l} \hline \hline
    {$M_\star$ ($M_\odot$)} & {$L_\star$ ($L_\odot$)} & {$\omega$} & {$R_\mathrm{p}$ ($R_\odot$)} & {$R_\mathrm{eq}$ ($R_\odot$)} & {$T_\mathrm{pole}$ (K)} & {$T_\mathrm{eq}$ (K)} \\ \hline
     1.35 & 3.657 & 0.117 & 1.482 & 1.485 & 6561 & 6544\\
     1.45 & 5.165 & 0.262 & 1.607 & 1.624 & 6887 & 6805\\ 
     1.55 & 6.938 & 0.328 & 1.709 & 1.740 & 7203 & 7073\\ 
     1.65 & 8.948 & 0.618 & 1.751 & 1.872 & 7688 & 7178\\ 
     1.775 & 11.987 & 0.635 & 1.815 & 1.949 & 8131 & 7574\\ 
     1.925 & 16.706 & 0.698 & 1.881 & 2.059 & 8715 & 7995\\
     2.2 & 28.585 & 0.850 & 1.907 & 2.236 & 10015 & 8703\\ \hline
\end{tabular}
\caption{Theoretical flattened star parameters for a range of masses. We obtain rotation rates from \citet{wolff1997angular}, which calculated average rotational velocities from hundreds of $v\sin(i)$ observations of A/F stars. Our chosen masses correspond to the mass bins in \citet{wolff1997angular}. We adopt spherical radii and temperatures from \citet{eker2018interrelated} and estimate each star's gravity-darkened radii and temperatures using the Roche model of a rotating star \citep{lara2011gravity}.}

\label{table:theoreticalstars}
\end{table*}

\subsubsection{Observed Rapid Rotators}\label{sec:observed}
We model the stellar output from five famous rapid rotators whose gravity-darkened bulk parameters have been observed interferometrically (Table \ref{table:famousstars}). We also include a sixth star in KELT-9 as a follow-up to our previous works on this system \citep{ahlers2020kelt,cauley2022effects}. We use measured masses, polar radii, polar temperatures, and rotation rates as inputs for our stellar model and calculate the star's radius at every colatitude via
\begin{equation}
    \frac{1}{\omega\tilde{r}}+\frac{1}{2}\sin^2(\theta)=\frac{1}{\omega^2}+\frac{1}{2},
\end{equation}
where $\omega$ is the star's fractional breakup rotation rate and $\tilde{r}\equiv R(\theta)/R_\mathrm{eq}$ is the star's normalized radius as a function of its colatitude $\theta$ \citep{lara2011gravity}. We then numerically solve for each star's surface temperature $T(\theta)$ as a function of colatitude using Equation 31 of \citet{lara2011gravity}.

Using $R(\theta)$ and $T(\theta)$, we generate SEDs using a two-step process. We first calculate spectra from 100 \AA\ to 2.0 $\mu$m using \texttt{Spectroscopy Made Easy (SME)} \citep{piskunov17} in steps of 100 K for a temperature grid spanning the polar to equatorial temperatures of the star. We compute the spectra at 15 $\mu$-angles from cos$\theta = \mu = 1$ to $\mu = 0.001$. We extract the spectral line lists and atomic parameters from VALD \citep{piskunov95} using the relevant stellar parameters. We then construct a 2D grid with cell size $0.01 R_\text{eq} \times 0.01 R_\text{eq}$ and interpolate the $R(\theta)$ and $T(\theta)$ values onto the Cartesian stellar surface. We loop over the stellar surface and generate the stellar spectrum at each grid point by interpolating the synthesized \texttt{SME} spectra onto that cell's specific temperature and $\mu$-angle. We then sum the intensities across the entire surface to produce the final stellar spectrum.

\begin{figure*}[htbp]
    \centering
    \includegraphics[width=\textwidth]{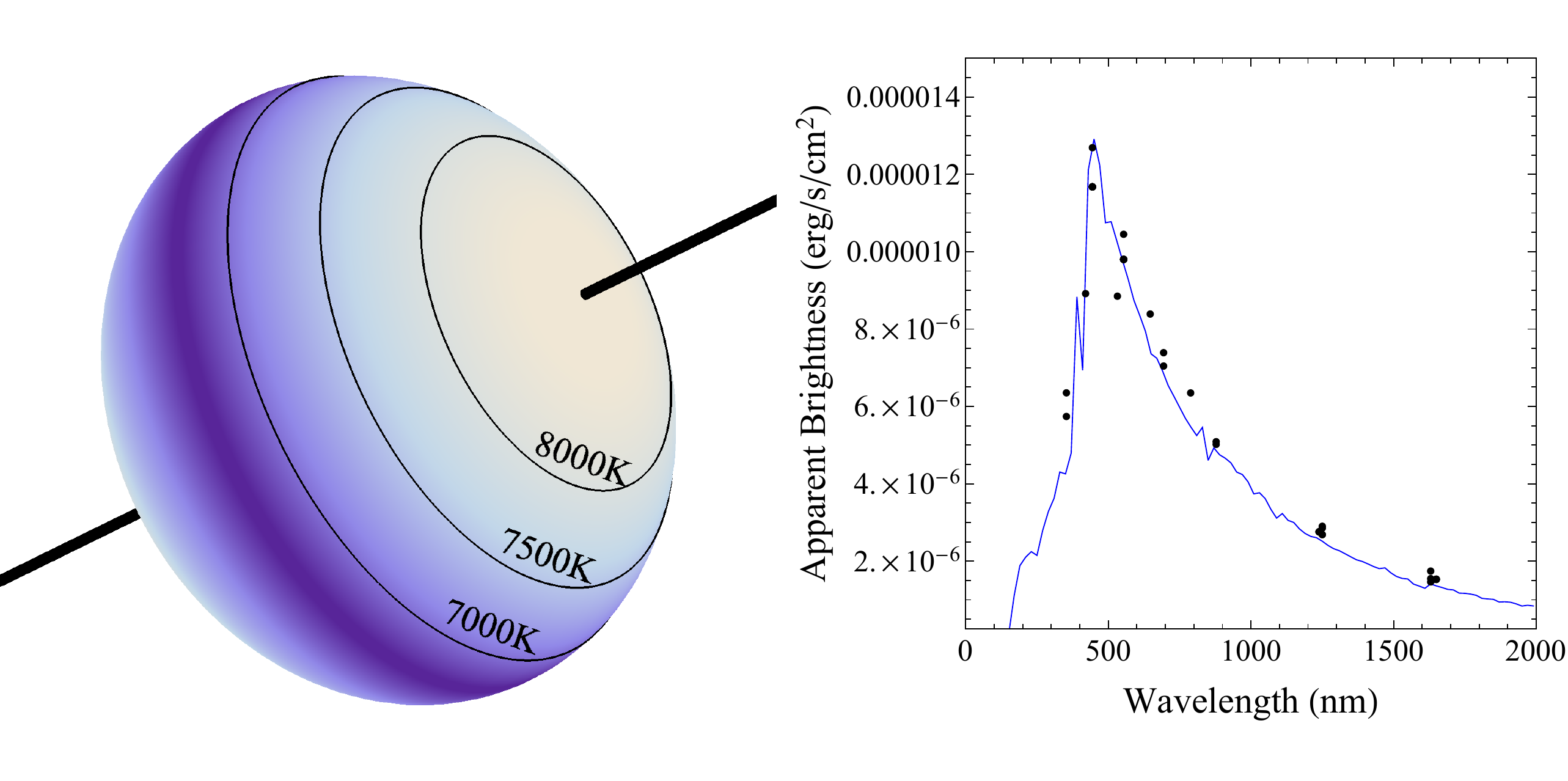}
    \caption{We test our gravity-darkened model of Altair against observations of its spectral energy distribution (SED). On the left, we model the gravity-darkening of Altair by accounting for both temperature changes on the surface and flattening caused by rapid rotation. From pole to equator, the temperature varies from 8450 K to 6823 K, and the stellar radius varies by $\sim24$\%. On the right, we model the SED for Altair and compare it to observation values at a stellar inclination of $57.2^\circ$ as seen from Earth. Observed data are from several works \citep{morel1978ubvrijklmnh,hindsley1994us,ofek2008calibrated,shenavrin2011search,gaspar2013collisional,huang2015empirical,soubiran2016pastel} and are taken from the VizieR database of astronomical catalogues \citep{ochsenbein2000vizier}}. We list Altair's gravity-darkened SED parameters in Table \ref{table:famousstars}.
    \label{fig:pov}
\end{figure*}


\subsubsection{Theoretical Rapid Rotators}
We also apply our gravity-darkened model to a range of theoretical stars, spanning from 1.35 $M_\odot$ to 2.2 $M_\odot$. We use average mass and rotation rates from \cite{wolff1997angular}, which observed hundreds of rapidly rotating stars. We first estimate each star's spherical radius and luminosity from the mass-radius relation and mass-luminosity relation, respectively \citep{eker2018interrelated}, and translate those spherical values into flattened parameters via the Roche model, following \citet{lara2011gravity}. We list relevant parameters in Table \ref{table:theoreticalstars}.

\subsection{Planet Instellation}\label{sec:irradiance}
Using our gravity-darkened SED model described in Section \ref{sec:observed}, we calculate instellations for hypothetical Earth-like planets orbiting the observed stars listed in Table \ref{table:famousstars} and theoretical stars listed in Table \ref{table:theoreticalstars}. Following \citet{ahlers2016gravity} and Wilson \& Ahlers (2021, submitted), we calculate instellation including rapid stellar rotation and planet spin-orbit misalignment. We use these planet instellation calculations as inputs for our 1D climate model.

\subsection{1D Climate Model}\label{sec:climatemodel}
We used a 1D radiative-convective, cloud-free climate model \citep{kopparapu2013habitable,kopparapu2014habitable} to calculate the inner and outer HZ edges of our selected stars. More details about the model are given in these papers and corresponding references. To simulate the effect of rotational flattening between the pole and the equator, we used the SED of the star at the equator and at the pole. These SEDs are used to estimate the incident stellar flux on a planet around stars shown in Table 2. The methodology to calculate the inner habitable zone and the outer habitable zone is similar to \citet{kasting1993habitable,kopparapu2013habitable,kopparapu2014habitable}. 

The inner edge of the HZ is estimated by assuming a fully saturated atmosphere on an Earth-size planet, and increasing the surface temperature from 220 K up to 2200 K. The effective stellar flux $S_{eff}$, which is the value of solar constant required to maintain a given surface temperature, is calculated from the ratio between the net outgoing IR flux and the net incident solar flux, both evaluated at the top of the atmosphere. When the model reaches an asymptotic $S_{eff}$ value, the atmosphere is optically thick to the outgoing IR radiation, and the planet enters the runaway greenhouse regime. This is considered to be the inner edge of the HZ and the corresponding value of $S_{eff}$ is noted as the inner HZ limit. The total flux incident at the top of the atmosphere is taken to be the present solar constant at Earth's orbit 1360 W. m$^{-2}$. The model then calculates how out of energy balance the planet is to find the system's inner HZ limit. See \citet{kopparapu2013habitable} and \citet{kopparapu2014habitable} for more detail on this process.

The outer edge of the HZ is calculated by fixing the surface temperature of an Earth-like planet with a 1 bar N$_2$ atmosphere, and the atmospheric CO$_{2}$ partial pressure was varied from 1 to 35 bar. Due to competing effects of the outgoing IR and the incoming solar, $S_{eff}$ experiences a minimum as a function of CO$_{2}$ partial pressure. This minimum is where the ``maximum'' amount of warming can be achieved with CO$_{2}$, and this is the ``maximum greenhouse'' limit for the outer edge of the HZ. The corresponding $S_{eff}$ at this minimum value is noted as the outer HZ.

We calculate inner and outer HZ edges by allowing the planet to equilibrate to an instantaneous flux level at 200 locations throughout the planet's orbit. Figure \ref{fig:power} shows how a planet's irradation can change as it moves throughout its orbit. The rest of our results only show planet behavior when residing at the stellar equator or above a pole because those are the minimum and maximum fluxes, respectively, with all other locations along the orbit being intermediate scenarios.

\section{Results}\label{sec:results}
We find that rapid stellar rotation has a number of significant effects on an orbiting Earth-like planet. Rapid rotation commonly observed in stars $M_\star\geq1.3M_\odot$ distorts stars into oblate shapes and induces a pole-to-equator temperature gradient, which can impact an orbiting planet's irradiance.

To summarize, stellar gravity darkening caused by rapid rotation:
\begin{itemize}
    \item Significantly reduces a planet's UV irradiation.
    \item Can induce global seasons on a planet at twice the planet's orbit frequency.
    \item Moderately impacts the system's habitable zone location.
    \item Needs to be accounted for in exoplanet observations.
\end{itemize}

We detail each of these findings in the following subsections. 

\subsection{Reduced UV Irradiation}
Rapid stellar rotation produces two effects that can decrease a planet's irradiation. First, the star's rotation induces a flattened shape that changes the total area of its sky-projected disk. When viewing the star equator-on, the star's projected area is smaller than when viewing pole-on, decreasing the planet's total flux from that viewing angle. Second, stellar rotation produces a pole-to-equator temperature gradient that can cause the star's equator to be up to 2000 K cooler than its poles. The lower temperature causes the star to appear dimmer at the equator, which can also decrease a planet's irradiation. These two effects are additive: both stellar flattening and the temperature gradient cause the a planet to receive less total starlight, and their effects on irradiation are maximized when the planet resides in the stellar equatorial plane.

We find that rapid stellar rotation has the largest impact on a planet's UV irradiation. The star's temperature gradient mainly reduces output near the star's peak emission wavelength (An orbiting planet therefore receives less UV flux when residing near the host star's equatorial plane.

Figure \ref{fig:power} shows a hypothetical planet in a $90^\circ$ inclined orbit around Altair. The left plot shows the difference between stellar SEDs seen by the planet when residing in the star's equatorial plane ($t_1$) and when above the stellar pole ($t_2$). The difference in incident power density occurs mainly in the near UV and optical wavelengths. When in an inclined orbital configuration, a planet's incident flux cycles between periods of less/greater UV irradiation as its exposure to the star's hot poles varies. The amplitude of this variation depends on the strength of the stellar temperature gradient and on the planet's orbital inclination; in general, a more inclined orbit translates to larger variations in received flux.

\subsection{Gravity-Darkened Seasons}
We find that stellar gravity darkening can cause a planet's irradiation to vary by as much as 40\% when in a polar orbital configuration. When inclined, the planet's exposure to the star's bright poles and dim equator vary at a rate of twice the orbit frequency. In general, the more inclined the orbit, the more dramatic the planet's irradiation can vary. In an orbit coplanar with the star's equator, the planet receives instellation that is affected by gravity darkening, but is constant in time. In a $90^\circ$ inclined orbit, the planet's exposure to the hot stellar poles is maximized, causing dramatic changes in irradiance throughout the orbit. Following \citet{ahlers2016gravity}, we refer to this phenomenon as ``gravity-darkened seasons''. 

\begin{figure*}[htbp]
    \centering
    \includegraphics[width=1.0\textwidth]{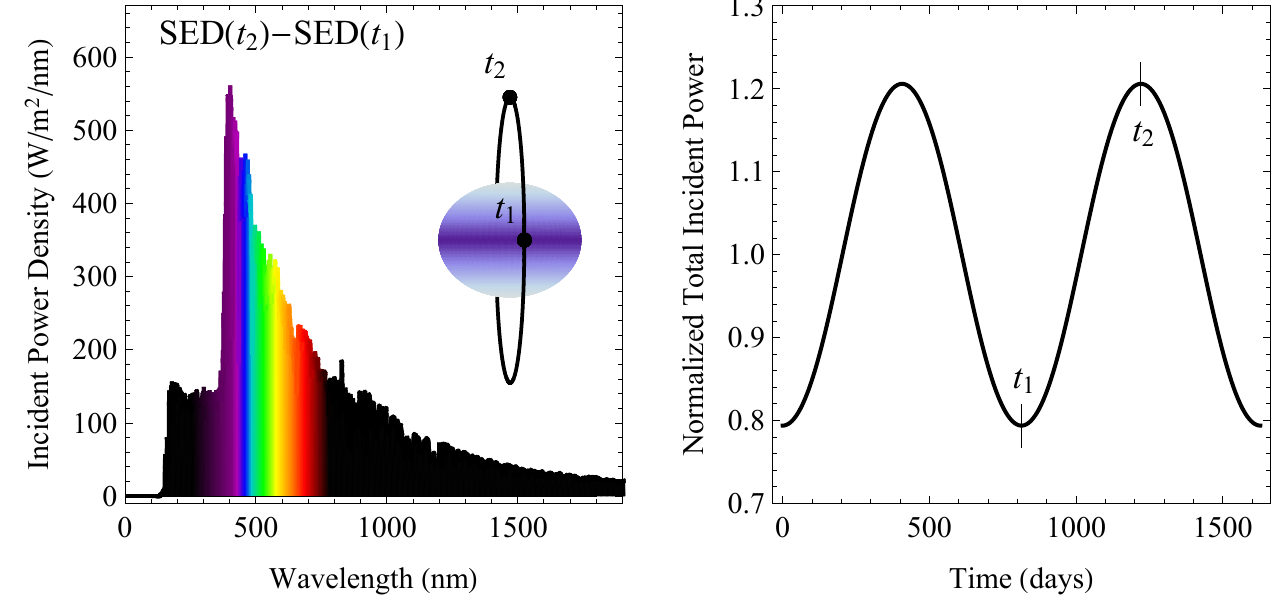}
    \caption{Irradiance of a planet in a $90^\circ$ inclined circular orbit around Altair at 3 AU. (Left) The difference in incident power density when the planet resides within the stellar equatorial plane and above the north stellar pole  ($t_1$ and $t_2$, respectively). As the planet varies in exposure between the star's hotter poles and cooler equator, its irradiance varies primarily in the UV, with almost no change in IR incident power. (Right) The planet's irradiance as a function of time. The total incident power on the planet varies by $\sim40$\% in this scenario -- exclusively due to the star's asymmetric brightness and shape. We calculate this curve by modeling Altair's apparent SED for 200 locations throughout the planet's polar orbit and measuring the total incident flux at each location.}
    \label{fig:power}
\end{figure*}

To demonstrate the dramatic effects that gravity-darkened seasons can have on a planet, we model a planet around Altair in a $90^\circ$ inclined orbit (Figure \ref{fig:power}). We place the planet at a distance of 3 AU, which corresponds to an average flux density similar to Earth ($\sim1360\mathrm{W/m^2}$). When moving throughout its orbit, the planet receives far more starlight when above the stellar poles and the least starlight when residing in the star's equatorial plane. The optical and near UV wavelengths vary the most since this is the wavelength region where the blackbody emission peaks for the temperatures across Altair's surface, i.e., the $T=6800$ K equatorial regions peak at $\approx 426$ nm and the hot $T=8500$ K polar regions peak at $\approx 341$ nm.

\subsection{Habitable Zone Location}
\begin{figure*}[htbp]
 \centering
 \includegraphics[width=0.9\textwidth]{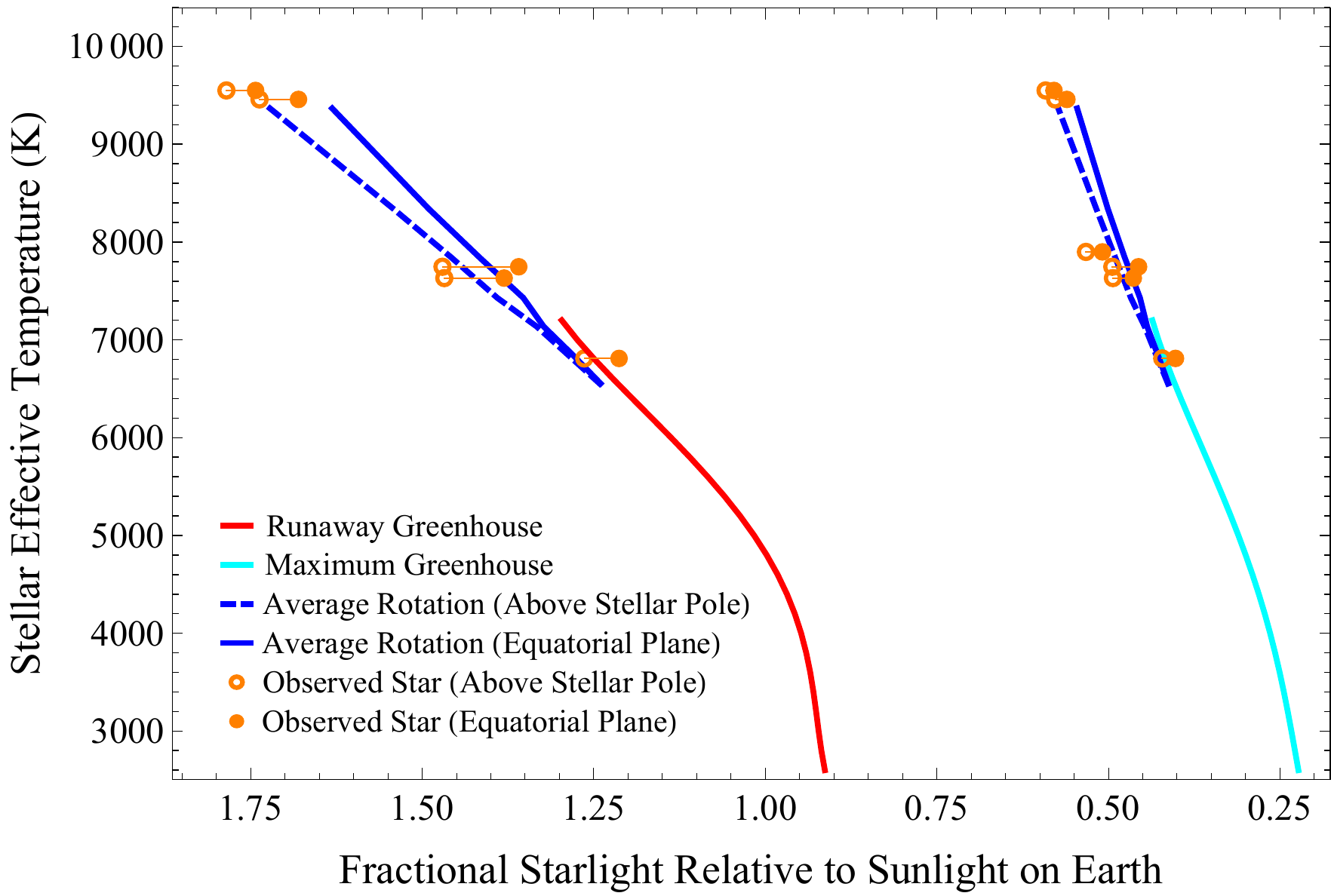}
 \caption{We extend the habitable zone limits calculated by \citet{kopparapu2014habitable} up to $T_\mathrm{eff}~10000$ K and find that stellar gravity darkening causes habitable zone locations to vary based on planet orbit geometry. The dark runaway greenhouse (red) and maximum greenhouse (cyan) boundaries are calculated using the 1D climate model from \citet{kopparapu2014habitable}. We extend calculations to span A/F stars and include irradiance deviations due to rapid stellar rotation. The blue lines indicate habitable zone boundaries calculated using average stellar rotation values as a function of mass (Table \ref{table:theoreticalstars}. The dashed blue lines correspond to HZ boundaries when the planet is residing in the star's equatorial plane, and the solid blue lines mark HZ boundaries when residing above a stellar pole. }
 \label{fig:habzone}
\end{figure*}

Stellar gravity darkening can directly impact the location of a system's habitable zone. Both the star's flattened shape and temperature gradient influence the total flux on an orbiting planet, which we show in Figure \ref{fig:power}. The pole-to-equator differences in the luminosity can influence the incident stellar flux on the planet, and the corresponding location of the inner and outer HZ. Fig. \ref{fig:habzone} shows the traditional HZs around main-sequence stars with no rotation (red for inner HZ limit or the runaway greenhouse limit, cyan for outer HZ limit or the maximum greenhouse limit). Plotted on top of them are HZs calculated in this study, when a planet is viewing the pole of the star (open orange circle), and when it is at the equator (filled orange circle). Consistent with previous studies, the trend is that as one moves to hotter stars the inner HZ {\it flux} boundary moves closer to the star. This is because hotter stars have their peak radiation shifted more towards the blue part of the electro-magnetic spectrum. As we included Rayleigh scattering by water vapor in our model (which is proportional to $1/\lambda^{4}$), and because the planet has a fully saturated water-vapor atmosphere, the corresponding planetary albedo increases which essentially means that to reach a runaway greenhouse limit (the inner HZ) for this high planetary albedo, the planet needs to receive more incident stellar flux around a hotter star. 

A similar reasoning could be applied to understand the differences in the HZ limits when the planet is around the pole and the equator.  Fig. \ref{fig:power}, left panel, shows the difference in power density from pole to equator. And the difference is more pronounced towards the shorter wavelengths, which indicates that there are more bluer wavelength photons reaching the planet near the pole than the equator. Consequently, the corresponding planetary albedo will be relatively higher near the poles, and hence a higher $S_{eff}$ value as can be seen by the open orange circles in Figure \ref{fig:habzone}. 

While the inner HZ differences for pole-to-equator shows noticeable differences, a similar comparison of outer HZ limit shows muted or no variation in this limit. This is because the outer edge of the HZ is calculated by increasing the CO$_{2}$ in the atmosphere, and then finding out at which value of CO$_{2}$ does the planet no longer warms (the maximum greenhouse limit). During this process, CO$_{2}$ condensation (ice) accumulates, which steadily increases the planetary albedo. However, CO$_{2}$ is also an effective greenhouse gas and absorbs the outgoing IR radiation and tries to warm the planet.  Eventually, as CO$_{2}$ increases, the atmosphere becomes optically thick at all IR wavelengths  and, at the same time, the Rayleigh scattering due to CO$_{2}$ condensation increases planetary albedo. The competing effects of these two, trying to increase and decrease the warming of the planet, results in essentially similar outer HZ limit as earlier results. 

\subsection{Impacts on Modeling Habitability}

To highlight the overall impact that stellar gravity darkening can have on modeling habitability, we test whether accounting for gravity darkening produces a noticeably different environment versus using a spherical star. This subsection is motivated by previous works that have studied planets around high-mass stars via transit photometry \citep[e.g.,][]{gaudi2017giant,talens2017mascara}, radial velocity \citep[e.g.,][]{galland2006extrasolar,lagrange2009extrasolar}, Doppler tomography \citep[e.g.,][]{zhou2016kelt,dorval2020mascara}, or habitability modeling \citep[e.g.,][]{ramirez2018methane} but have not taken the host star's gravity darkening into account. 

\begin{figure*}[htbp]
\centering
\includegraphics[width=1.0\textwidth]{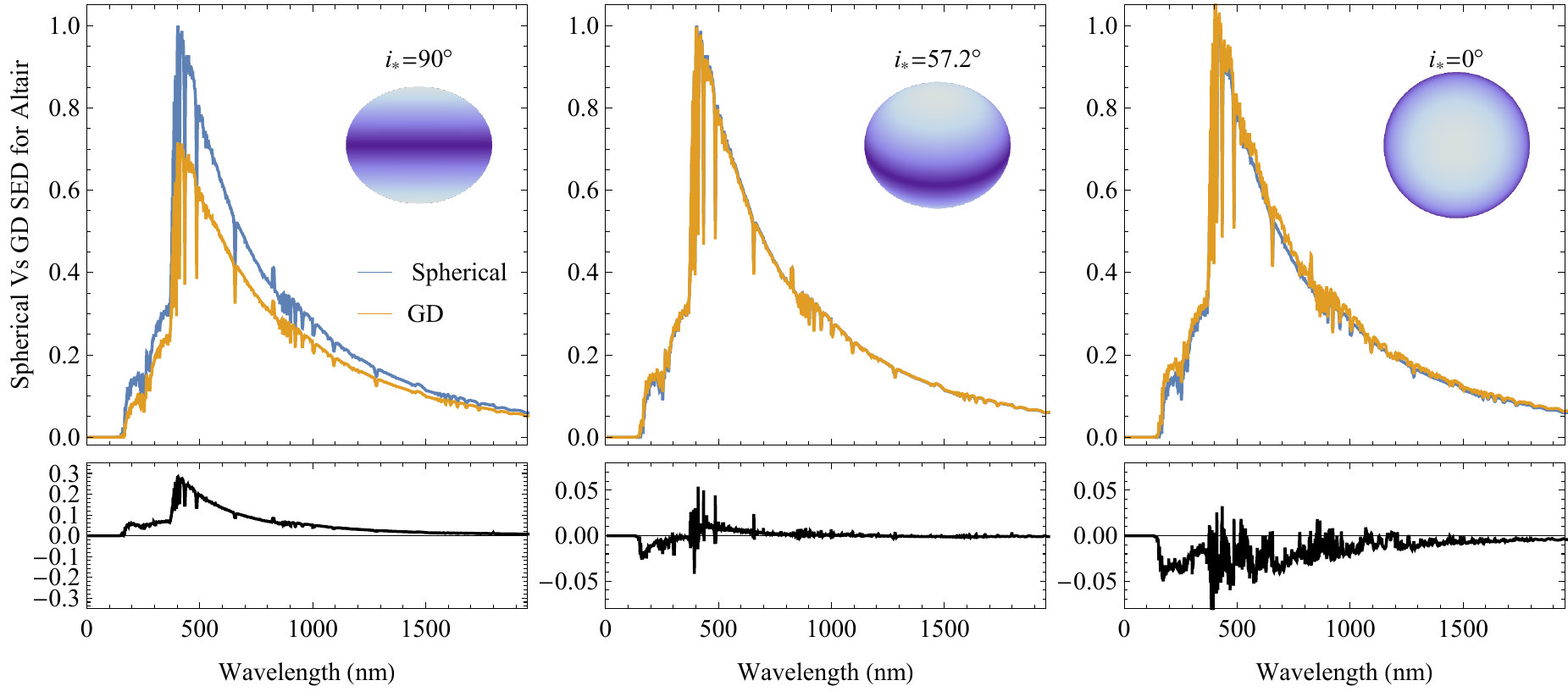}
\caption{Stellar gravity darkening can massively impact calculated spectral energy distributions of rapidly rotating stars. As a comparison, we model Altair's SED with and without gravity darkening included. For the spherical (blue) case, we use approximated values of Altair from \citet{turnbull2015exocat} and listed on the NASA Exoplanet archive}. Our gravity-darkened model (yellow) uses values from \cite{monnier2007imaging}. We compare the spherical model to the gravity-darkened model at three viewing geometries: $i_\star=90^\circ$ (equator-on), $i_\star=57.2^\circ$ (Altair's stellar inclination as seen from Earth), and $i_\star=0^\circ$ (pole-on) . The lower plots show the difference between the two models for each scenario.
\label{fig:compare}
\end{figure*}

We consider how a spherical versus gravity-darkened stellar model can yield different spectral energy distributions for a hypothetical planet around Altair. Imagine a planet were discovered orbiting within Altair's habitable zone. A common starting point toward estimating the planet's overall habitability would be to model its irradiance and equilibrium temperature based on stellar values listed in exoplanet repositories such as the Exoplanet Archive website or the \emph{TESS} input catalog (TIC) \citep{stassun2019revised}. The NASA Exoplanet Archive currently lists the following values for Altair: $T_\mathrm{eff}=7800$ K, $\log(g)=3.76~(\log_{10}(\mathrm{cm/s^2}))$, $R_\star=1.79~R_\odot$, $M_\star=1.83~M_\odot$. These values come from a bulk survey of 2347 stars \citep{turnbull2015exocat}, where estimates of stellar temperature, radius, and mass are calculated via analytic expressions based on observed $B-V$ color and luminosity. This survey provides useful approximations to stellar properties, but makes no attempt to account for stellar rotation or gravity darkening.

We treat Altair's values listed on the NASA Exoplanet Archive as a ``spherical approximation'' as a comparison to Altair's gravity-darkened properties measured by \citet{monnier2007imaging}. We use both sets of parameters as inputs for a spherical approximation and gravity-darkened model, respectively, and calculate the SED that a planet would see at different orbit geometries. We show the SEDs in Figure \ref{fig:compare}.

The gravity-darkened and spherical approximation models differ in a key way: the gravity-darkened star's SED varies as a function of viewing geometry, whereas the spherical approximation SED is consistent at all viewpoints. In Figure \ref{fig:compare}, we compare the spherical model to three different viewing geometries: $i_\star=90^\circ$ (equator-on), $i_\star=57.2^\circ$ (Altair's inclination seen from Earth), and $i_\star=0^\circ$ (pole-on). The gravity-darkened SEDs vary dramatically due to the star's temperature gradient and oblate shape, but the spherical approximation SEDs are constant.

The values from the NASA Exoplanet Archive that we used for the spherical model are based on observations from Earth-- i.e., they specifically approximate Altair's SED at $i_\star=57.2^\circ$ viewing geometry. For that reason, the middle panel of Figure \ref{fig:compare} shows the best overall agreement between the spherical and gravity-darkened models. If Altair's $i_\star$ relative to Earth were a different value, its measured parameters listed in the Exoplanet Archive would be different-- as would its calculated SED.

Therefore, using spherical values for a gravity darkened star can significantly change estimated bulk parameters of an orbiting planet. The planet's equilibrium temperature will likely be inaccurate because the star's spherical $T_\mathrm{eff}$ approximation improperly represents the range of temperatures on the star's gravity-darkened surface. Moreover, the starlight that the planet receives is potentially drastically different that the starlight we observe. If the hypothetical planet around Altair is spin-orbit aligned (i.e., resides in the stellar equatorial plane at all times), then it would receive a constant instellation corresponding to the gravity-darkened scenario in panel 1 of Figure \ref{fig:compare}. If gravity darkening is not properly accounted for, one would estimate the planet's SED based on the Spherical scenario in Figure \ref{fig:compare} and overestimate the planet's equilibrium temperature and received starlight -- particularly in the optical and UV.

Observations of gravity-darkened host stars almost never match the received starlight of an orbiting planet. The only case where observation matches planet instellation throughout the planet's orbit is at equator-on viewing geometry ($i_\star=90^\circ$) and the planet is spin-orbit aligned. All other system configurations ($i_\star\neq90^\circ$ or spin-orbit angle is nonzero) guarantee a disconnect between observation and planet instellation.

\section{Discussion}\label{sec:discussion}
In this work, we show that rapid stellar rotation commonly exhibited by massive stars ($M_\star\gtrsim1.3M_\odot$) can dramatically impact the overall habitability of orbiting planets. Stellar gravity darkening and flattening due to rotation change a planet's total received starlight, which shifts the system's habitable zone. We discuss in the following subsections whether stars in this mass range can host habitable worlds, how this work relates to ongoing and future exoplanet missions, and what steps can be taken in the future to better characterize this phenomenon.

\subsection{Can A/F-type Stars Host Habitable Worlds?}
A common argument put forward in the literature against high-mass stars as habitable worlds is that the emit too much UV light for orbiting planets to be considered habitable \citep[e.g.,][]{buccino2007uv,guo2010habitable}. However, the results in this work indicate that planets in orbit around rapidly rotating high-mass stars would experience a reduction in the magnitude of UV stellar irradiation when compared to an equivalent non-rotating and non-oblate star. The reduction in UV irradiation is a result of the oblateness of the star, which causes a reduction in total flux received by the planet, as well as a pole-to-equator temperature gradient induced by rotation. Planets in an equatorial orbit around a rapidly rotating high-mass host star would experience the greatest reduction in UV irradiation, which may increase the potential for such planets to develop life. Planets in an inclined orbit would experience periodic changes in the stellar UV flux, which could pose challenges to the development of life when the UV flux is high, although the cyclical patterns of such a UV flux might also provide some benefits to the evolution of life. In general, the reduction of UV irradiation on planets orbiting rapidly rotating high-mass stars suggests that such planets cannot be excluded from being habitable based solely on stellar type.

\subsection{Relevance to Ongoing and Future Missions}
A/F stars make up a large fraction of currently-discovered exoplanet hosts. As of January 1, 2022, nearly 10\% of confirmed planets (443 of 4569) orbit host stars with $M_\star\geq1.3M_\star$.\footnote{\href{https://doi.org/10.26133/NEA12}{https://doi.org/10.26133/NEA12}} Over 18\% of \emph{TESS} Targets of Interest (869 of 4704) orbit stars in this mass range\footnote{\href{https://doi.org/10.26134/exofop3 }{https://doi.org/10.26134/exofop3}}. Ostensibly, all of these planets orbit rapid rotators. \citet{barclay2018revised} estimates that \emph{TESS} will find as many as 2000 total planets orbiting A/F stars, comprising over 40\% of \emph{TESS}'s total exoplanet yield. NASA's upcoming Nancy Grace Roman telescope, which is expected to find 100,000 transiting exoplanets, will likely find many thousand planets orbiting rapid rotators as well \citep{montet2017measuring}.

Rapid stellar rotation needs to be accounted for when characterizing these planets. Previous papers have demonstrated how gravity darkening can impact transit photometry due to the star's asymmetry \citep[e.g.,][]{ahlers2019dealing,ahlers2020gravity,ahlers2020kelt}. \citet{wong2020exploring} found evidence of gravity darkening impacting phase curve analysis of hot Jupiter KELT-9b by causing its upper atmosphere to vary in temperature, and Wilson \& Ahlers (2021, submitted) demonstrated that gravity darkening can slightly skew transmission spectroscopy observations. Accounting for rapid stellar rotation allows for better measurements of orbiting planets' bulk properties and reduces systematics in atmospheric characterization.

The relevance of rapid stellar rotation will continue to grow as future missions expand our ability to explore planets around A/F stars. A-type stars are particularly under-explored, with nearly 40\% of confirmed planets (91 of 239) being hot Jupiters with orbital periods shorter than 10 days. As ongoing all-sky surveys such as \emph{TESS} or the ground-based Kilodegree Extremely Little Telescope will continue to find longer-period planets as their searches mature. The Nancy Grace Roman Telescope is planned for 72-day observing windows at high precision, which will undoubtedly find smaller, cooler planets around A/F stars. Additionally, the Decadal Survey on Astronomy and Astrophysics 2020 recently recommended a six-meter UV/Optical/IR space telescope as NASA's next flagship mission; such an instrument would allow for detailed measurements of how gravity darkening can affect planet atmospheres. High-precision NUV and optical spectroscopy would be particularly useful for understanding gravity-darkened seasons because variations in flux occur primarily in these wavelength ranges. Observation of a planet's atmosphere when undergoing gravity-darkened seasons would help reveal both how a planet responds to a strongly varying UV flux and, more generally, whether planets around gravity-darkened stars receive a favorable amount of UV light to be considered habitable.

\subsection{Future Work and Conclusions}
The phenomenon of gravity-darkened seasons is still largely unexplored. This work determines the impact of how gravity darkening can impact the location of the habitable zone and provides a first look as to how it can alter a planet's instellation; future work can investigate the spatially- and time-varying changes to planetary climate. Latitudinal energy balance models (EBMs; e.g., \citet{north2017energy}) could improve upon the 1D radiative convective model calculations in this work by evaluating the geographical effects of gravity darkening on a planet's temperature distribution. Gravity-darkened planets in an inclined orbit would experience changes in the total stellar flux at different locations in orbit, with the poles of the planet experiencing prolonged periods of darkness and light for more extreme inclinations. The ice-albedo feedback included in EBMs makes them well-suited for understanding the region of orbital space where a gravity-darkened planet would be expected to be warm and temperate versus completely frozen.

Future modeling can also demonstrate how the varying instellation due to gravity darkening can combine with traditional seasons. \citet{ahlers2016gravity} first showed that planets with tilted rotation axes can receive wildly varying hemispherical seasonal changes depending on their precession angle, including scenarios with extreme long winters or multiple summers per year. EBMs could be used to examine the combined impact of traditional and gravity-darkened seasons on the latitudinal temperature distribution over the course of an orbit. These scenarios also can be further explored with a global climate model (GCM) to understand the three-dimensional changes that would occur to a planet's surface and its atmosphere from gravity-darkened seasons. The model development required for such scenarios is much greater for a GCM than an EBM, but ultimately a hierarchy of modeling approaches will be needed to understand the impact of gravity-darkening on climate. We currently know that the combination of gravity-darkened host stars and spin-orbit misaligned planets is commonplace in these systems and is impactful on planet instellation. However, future works can reveal how planets behave in such dynamic environments and can better determine whether high-mass stars should be considered in the search for habitable worlds.

\acknowledgements
J.P.A.’s research was supported by an appointment to the NASA Postdoctoral Program at the NASA Goddard Space Flight center, administered by Universities Space Research Association under contract with NASA. J.P.A. acknowledges support from the GSFC Sellers Exoplanet Environments Collaboration (SEEC), which is funded in part by the NASA Planetary Science Division’s Internal Scientist Funding Model. J.H.M. acknowledges funding from the NASA Habitable Worlds program under award 80NSSC20K0230. This work has made use of the VALD database, operated at Uppsala University, the Institute of Astronomy RAS in Moscow, and the University of Vienna. This research has made use of the NASA Exoplanet Archive, which is operated by the California Institute of Technology, under contract with the National Aeronautics and Space Administration under the Exoplanet Exploration Program.

\bibliography{citations}
\bibliographystyle{aasjournal}
\end{document}